\documentclass[%
 reprint,
 amsmath,amssymb,
 aps,
]{revtex4-2}

\usepackage{color}
\definecolor{RedWine}{rgb}{0.743,0,0}
\definecolor{green(pigment)}{rgb}{0.0,0.65,0.31}
\definecolor{orange}{rgb}{0.901,0.475, 0.0} 

\newcommand{\knl}{\text{$k_{n\ell}$}}
\newcommand{\knlp}{\text{$k_{n'\ell}$}}
\newcommand{\fs}{f\sigma_8}
\newcommand{\bs}{b\sigma_8}
\newcommand{\wnl}{\mathcal{W}_{n\ell}}
\newcommand{\fnl}{f_\text{NL}}

\newcommand{\rmin}{{r_{\text{min}}}}
\newcommand{\rmax}{{r_{\text{max}}}}
\newcommand{\nobs}{{N_{\ell n n'}^{\text{obs}}}}
\newcommand{\fsky}{f_\mathrm{sky}}

\usepackage{siunitx}
\usepackage{aas_macros}
\usepackage{graphicx}
\usepackage{dcolumn}
\usepackage{bm}
\usepackage{ulem}  
\usepackage{booktabs}
\usepackage{hyperref}
\hypersetup{
    colorlinks = true
}
\usepackage[capitalise]{cleveref}

\begin{document}
\preprint{APS/123-QED}

\title{Fast Theoretical Predictions for Spherical Fourier Analysis of Large-Scale Structures}


\author{Brandon Khek}
\affiliation{Rice University, Houston, TX, 77005, USA}
\affiliation{University of Pennsylvania, Philadelphia, PA, 19104, USA}

\author{Henry Grasshorn~Gebhardt}
\affiliation{Jet Propulsion Laboratory, California Institute of Technology, Pasadena, CA 91109, USA}
\affiliation{California Institute of Technology, Pasadena, CA 91125, USA}

\author{Olivier Dor{\'e}}
\affiliation{Jet Propulsion Laboratory, California Institute of Technology, Pasadena, CA 91109, USA}
\affiliation{California Institute of Technology, Pasadena, CA 91125, USA}


\begin{abstract}
    On-going or soon to come cosmological large-scale structure surveys such as DESI, SPHEREx, Euclid, or the High-Latitude Spectroscopic Survey of the Nancy Grace Roman Space Telescope promise unprecedented measurement of the clustering of galaxies on large scales. When quantified with the Cartesian Fourier basis, the measurement of these large scales requires the introduction of so-called wide-angle corrections. By contrast, the measurement of the power spectrum  in a spherical Fourier Bessel (SFB) basis does not require such corrections and naturally accounts for the spherical survey geometries. Here, we develop and implement a fast code to construct the SFB power spectrum and investigate how line of sight effects, physics such as non-Gaussianity, and differing survey geometries affect SFB power spectrum estimates. We then leverage our program to predict the tightness of cosmological constraints from realistic survey specifications using a Fisher matrix formalism.
\end{abstract}

\maketitle


\section{\label{sec:intro}Introduction} 

In order to answer questions about elusive cosmological phenomena such as dark energy and inflation, next-generation deep, wide-angle galaxy redshift surveys such as SPHEREx \footnote{\url{https://spherex.caltech.edu/}}, Euclid \footnote{\url{https://www.euclid-ec.org/}}, DESI \cite{desi}, and Roman \footnote{\url{https://roman.gsfc.nasa.gov/}} will be launched within the next decade and will obtain spectra of hundreds of millions of galaxies. This will enable the analysis of the large-scale distribution of matter throughout space, and as a consequence, precise estimates of the cosmological parameters that govern the evolution of the universe.

A three-dimensional map of the distribution of galaxies and galaxy clusters can be studied using statistics such as the two-point correlation function $\xi(\mathbf{x})$ which describes the probability of observing two galaxies separated by some distance. However, for separation of scales, in this paper we use the power spectrum, the two-point correlation function in Fourier space. The power spectrum contains the same information as $\xi(\mathbf{x})$, though has advantages in our case, such as the fact that the Fourier modes that comprise the power spectrum have the desirable property of being statistically orthogonal in a homogeneous and isotropic universe \cite[e.g.,][]{Peebles:1973ApJ...185..413P,Hamilton1998,GrasshornGebhardt+:2022JCAP...01..038G}.
Further details on trade-offs between the two-point function in configuration space and Fourier space can be found in \citet{feldman}.

Previous measurements of the 3D power spectrum have relied on Cartesian Fourier decomposition and have been sufficient for pencil-beam surveys with small volumes and minimal sky coverage. In the global-plane parallel case, one approximates the sky as being planar, and the Cartesian power spectrum uses a single global line of sight (LOS) for each galaxy pair. However, this assumption neglects the spherical geometry of the sky and becomes invalid for galaxies with large angular separations as is expected in upcoming wide-angle surveys. Resolutions include the local plane-parallel approximation, which uses a single LOS per galaxy pair \cite{yamamoto} and usually involves perturbative wide-angle corrections \citep{castorina,Beutler+:2019JCAP...03..040B,Benabou+Sands+:2024arXiv240404811B} (but see \cite{Castorina+:2022JCAP...01..061C,Wen+:2024arXiv240404812W} for nonperturbative approaches). Still, these perturbative approximations break down with the full-sky coverage to be seen in next-generation surveys, resulting in a loss of information. The mismatch of the Cartesian basis and the inherent spherical geometry of the survey implies that $P(\mathbf{k})$ is not directly measurable due to these wide angle effects \cite{Wang_2020}. While large angular scales pose many challenges, large radial scales also present difficulties in a Cartesian basis. For example, the survey volume is divided into redshift bins, and the power spectrum estimator is measured at the effective redshift in each bin. Radial bins that are too large inadequately capture redshift evolution while narrow bins fail to measure radial modes larger than the bin size \cite{Wang_2020}.

A spherical Fourier Bessel (SFB) decomposition of the matter overdensity field improves upon the Cartesian power spectrum with these limitations in mind. The basis functions of the SFB decomposition are defined as the eigenfunctions of the Laplacian operator in spherical coordinates: spherical Bessel functions $j_\ell$ and spherical harmonics $Y_{\ell m}$ for the radial and angular components of the Laplacian, respectively. In the SFB basis, a LOS for each galaxy is implicit, avoiding the need for flat-sky approximations, and we need only to assume that we can model redshift space distortions (RSD) to linear order \cite{Heavens:1994iq}. Furthermore, radial distortions are naturally included in the analysis and preserve isotropy, therefore maintaining statistical orthogonality of the angular modes in the real-space power spectrum \cite{Fisher_sfb}.

In this paper we develop code that computes the discrete SFB power spectrum, tunable to survey parameters such as the selection function and survey radius, as well as different cosmologies and physical parameters like galaxy bias. We use the code to study physical effects in the power spectrum -- including non-Gaussianity and RSD -- and investigate the evolution of power across multipole moments. We then leverage the power spectrum estimator to predict cosmological constraints with Fisher forecasts.

In \cref{sec:d-sfb}, we provide a short review of the SFB decomposition formalism along with the features obtained from various physical effects and survey parameters considered in our modeling. In \cref{sec:fisher}, for a SPHEREx-like survey, we predict constraints on the parameters $f\sigma_8,\,b\sigma_8, \, h,\,\Omega_m,$ and $\fnl$ using one galaxy number density and bias subsample, then extend the calculation to include cross-correlations with four other subsamples defined by their redshift uncertainties. We then conclude in \cref{sec:conclusions}. Our strategy for numerical integration is outlined in \cref{appendix:gaussleg}. Unless otherwise stated, we will use the \textit{Planck} $\Lambda$CDM cosmological model (2018) throughout this work.

\section{\label{sec:d-sfb} SFB Power Spectrum}
In this section, we will review the mathematical formalism of the SFB decomposition in both the continuous and discrete SFB bases, and demonstrate how physical effects, in particular galaxy number density evolution, redshift space distortions (RSD), and non-Gaussianity affect the shape and amplitude of the SFB power spectrum. This section concludes with a discussion of the shot noise computation we use in our analyses. In this work, we consider a survey with an observed comoving number density and galaxy bias corresponding to five SPHEREx redshift accuracy bins \cite{Dore:2014cca,Stickley:2016jwu,spherexpublicproducts,Feder+:2023arXiv231204636F}. In this section, unless otherwise noted, all plots of the SFB power spectrum and shot noise will be produced from the sample with the lowest redshift uncertainty $(\sigma(z)/(1+z)\leq 0.003)$ using a selection function that is a cubic spline of the values given in Ref.~\cite{spherexpublicproducts} and bias model obtained from linear regression from \cite{spherexpublicproducts}. 

Furthermore, we use a radial survey range of z $\in \left[0.0,\,4.6\right]$, and we generate the linear matter power spectrum $P(\mathbf{k})$ with \texttt{CAMB} \footnote{\url{https://camb.readthedocs.io/en/latest/}}. We will use these parameters for power spectrum estimates throughout the rest of this work. Finally, we only consider a spherically symmetric galaxy survey and leave a separate treatment of the angular mask for future work, which could be done using the coupling matrix as described in \cite{superfab}.

\textit{Continuous SFB Power Spectrum.} The SFB power spectrum is given by the covariance between continuous real space and SFB density contrasts which are defined by
\begin{align}
    \delta(\boldsymbol{r}) &= \int dk \sum_{\ell \,m} \sqrt{\frac{2}{\pi}}k\, j_\ell(kr)Y_{\ell\,m}(\theta, \, \phi)\,\delta_{\ell \, m}(k)\label{eq:deltar}
    \,,\\
    \delta_{\ell m}(k) &= \int d^3r  \left[\sqrt{\frac{2}{\pi}}k\, j_\ell(kr)Y_{\ell\,m}^*(\theta, \, \phi)\right]\delta(\boldsymbol{r}).
\end{align}
Following \cite{superfab}, we define an integral kernel
\begin{align}
    \label{eq:wl}
    \mathcal{W}_\ell(k,q) &= \frac{2qk}{\pi}\int dr\,\phi(r) D(r) b(r,q) j_\ell(kr)
    \,e^{-\frac12\sigma_{u+z}^2 q^2}
    \nonumber \\
    &\quad\times \sum_{\Delta\ell}(\delta^K_{\Delta\ell, 0} - \beta f^\ell_{\Delta \ell})j_{\ell + \Delta \ell}(qr)\,r^2
\end{align}
for radial selection function $\phi(r)$, linear growth factor $D(r)$, RSD parameter $\beta = f/b$, and possibly scale-dependent galaxy bias $b(r,q)$. The exponential factor models the \emph{Fingers-of-God} (FoG) effect as well as redshift errors by $\sigma^2_{u+z}=\sigma_u^2+\sigma_z^2$, where
\begin{align}
\sigma_u(r)^2 &= \frac{f(r)^2D(r)^2}{3}\int\frac{d^3k}{(2\pi)^3}\,\frac{P(k)}{k^2}
\end{align}
models the FoG velocity dispersion and $\sigma_z^2$ is the redshift-dependent redshift measurement uncertainty of the galaxy sample. As written, the exponential in \cref{eq:wl} is an approximation to $e^{-\frac12 \sigma^2\partial_{r}^2}j_\ell(qr)$. This approximation becomes inaccurate at large $\ell$ where it overestimates the suppression due to FoG. $P(k)$ is the matter power spectrum at $z=0$, and we model the evolution of density perturbations with the linear growth factor $D(r)$ and linear growth rate $f(r)$.
The sum in \cref{eq:wl} encodes the linear RSD contribution \cite{kaiser_rsd}, and the only non-zero $f^\ell_{\Delta \ell}$
\begin{align}
    f_{-2}^\ell &= \frac{\ell(\ell-1)}{(2\ell-1)(2\ell+1)}\,,\\
    f_{0}^\ell &= -\frac{2\ell^2+ 2\ell -1}{(2\ell-1)(2\ell+3)}\,,\\
    f_{2}^\ell &= \frac{(\ell+1)(\ell+2)}{(2\ell+1)(2\ell+3)},
\end{align}
arise from a recursion relation of the second derivative of the spherical Bessel function. The integral kernel captures the physics that ultimately serves to write the redshift-space, i.e. observed, galaxy density contrast to linear order in the SFB basis $\delta_{\ell m}^{\text{obs}}(k)$,
\begin{align}
    \label{eq:delta_obs_klm}
    \delta_{\ell m}^{\text{obs}}(k)= \int dq \,\mathcal{W}_\ell(k,q) \,\delta_{\ell m}(q)
    \,,
\end{align}
whereas $\delta_{\ell m}(q)$ denotes the density contrast of the matter field only.
We can then write the continuous SFB power spectrum as
\begin{align}
    C_\ell(k,k') &\equiv \langle \delta_{\ell m}^{\text{obs}}(k) \,\delta_{\ell m}^{\text{obs,*}}(k')\rangle \\
    &=\int dq\, \mathcal{W}_\ell(k,q)     \mathcal{W}^*_\ell(k',q) P(q)\,.\nonumber    \label{eq:cl_cont}
\end{align}

\textit{Discretized SFB Power Spectrum.} Boundary conditions must be imposed on the matter overdensity field in order to maintain orthogonality of the spherical Laplacian eigenfunctions when considering a finite survey radius from $\rmin$ to $\rmax$, resulting in a discretization of the SFB modes \cite{Fisher_BC}.

The discretized real space and SFB pair are defined by \cite{Samushia:2019arXiv190605866S,superfab}
\begin{align}
    \delta(\boldsymbol{r}) &= \sum_{n \ell m} \left[g_{n\ell}(r)Y_{\ell m}(\theta, \, \phi)\right]\delta_{n\ell m},\\
    \delta_{n\ell m} &= \int \int_\rmin^\rmax  r^2\, dr\,d\Omega\,g_{n\ell}(r)\,Y_{\ell m}^*(\theta, \, \phi)\,\delta(\boldsymbol{ r})\,, \label{eq:deltanlm}
\end{align}
where $g_{n\ell}$ is a linear combination of spherical Bessel functions of the first and second kind 
\begin{align}
    g_{n\ell}(r) = c_{n\ell} \,j_\ell(k_{n\ell}\,r) + d_{n\ell} \,y_\ell(k_{n\ell}\,r)
\end{align}
at discrete radial scales $k_{n\ell}$. The $g_{n\ell}$ serve as eigenfunctions to the  radial component of the Laplacian under potential boundary conditions with eigenvalues $-k_{n\ell}^2$, and in practice they are computed using the code \texttt{SphericalFourierBesselDecompositions.jl} \cite{superfab}.

Substituting \cref{eq:deltar} with the observed density contrast
into \cref{eq:deltanlm}, we exploit the orthogonality relation of spherical harmonics to obtain the relation between the continuous and discrete SFB-space density contrasts,
\begin{align}
    \delta_{n\ell m}^{\text{obs}}
    = \int
    \left[\int_{r_{\text{min}}}^{r_\text{max}}  g_{n\ell}(r)  \sqrt{\frac{2}{\pi}}k\, j_\ell(kr)
     \,r^2\, dr
    \right]
    \delta_{\ell \, m}^{\text{obs}}(k) \, dk\,. \label{eq:deltanlmgnl}
\end{align}
Modeling the observed density contrast as in \cref{eq:delta_obs_klm,eq:wl}, we obtain
\begin{align}
    \delta^\mathrm{obs}_{n\ell m}
    &= \int dq\,\mathcal{W}_{n \ell}(q) \, \delta_{\ell m}(q)
    \,,
\end{align}
where
\begin{align}
    \mathcal{W}_{n \ell}(q) &= \sqrt{\frac{2q^2}{\pi}}\int_{\rmin}^{\rmax}  g_{n\ell}(r) \, \phi(r) D(r) b(r,q)\,e^{-\frac12\sigma_{u+z}^2 q^2} \nonumber \\
     &\quad\times \sum_{\Delta\ell}(\delta^K_{\Delta\ell, 0} - \beta f^\ell_{\Delta \ell})j_{\ell + \Delta \ell}(qr)\, r^2 dr\,.\label{eq:wnl}
\end{align}
Compared to the continuous limit \cref{eq:wl}, in the discrete case, the first spherical Bessel is substituted by the discrete basis functions $g_{n\ell}(r)$ in \cref{eq:wnl}.
Thus, we define the discrete SFB power spectrum assuming isotropy around the observer ($\ell=\ell'$, $m=m'$) as the covariance between observed SFB modes $\delta^\mathrm{obs}_{n\ell m}$, or
\begin{align}
    C_\ell\left(\knl, k_{n'\ell}\right) &\equiv \langle \delta_{n\ell m}^\mathrm{obs} \, \delta_{n'\ell m}^\mathrm{obs} \rangle \nonumber \\
    &=   \int  \mathcal{W}_{n \ell}(q)  \mathcal{W}_{n' \ell}(q)P(q) \,dq.
    \label{eq:cl}
\end{align}
Obtaining this quantity in practice can be computationally intensive. For accurate results, we employ the procedure described in \cref{appendix:gaussleg}. 

\subsection{Radial selection function \label{sec:sel_functn}}
\begin{figure}
    \centering
    \includegraphics[width=.475\textwidth]{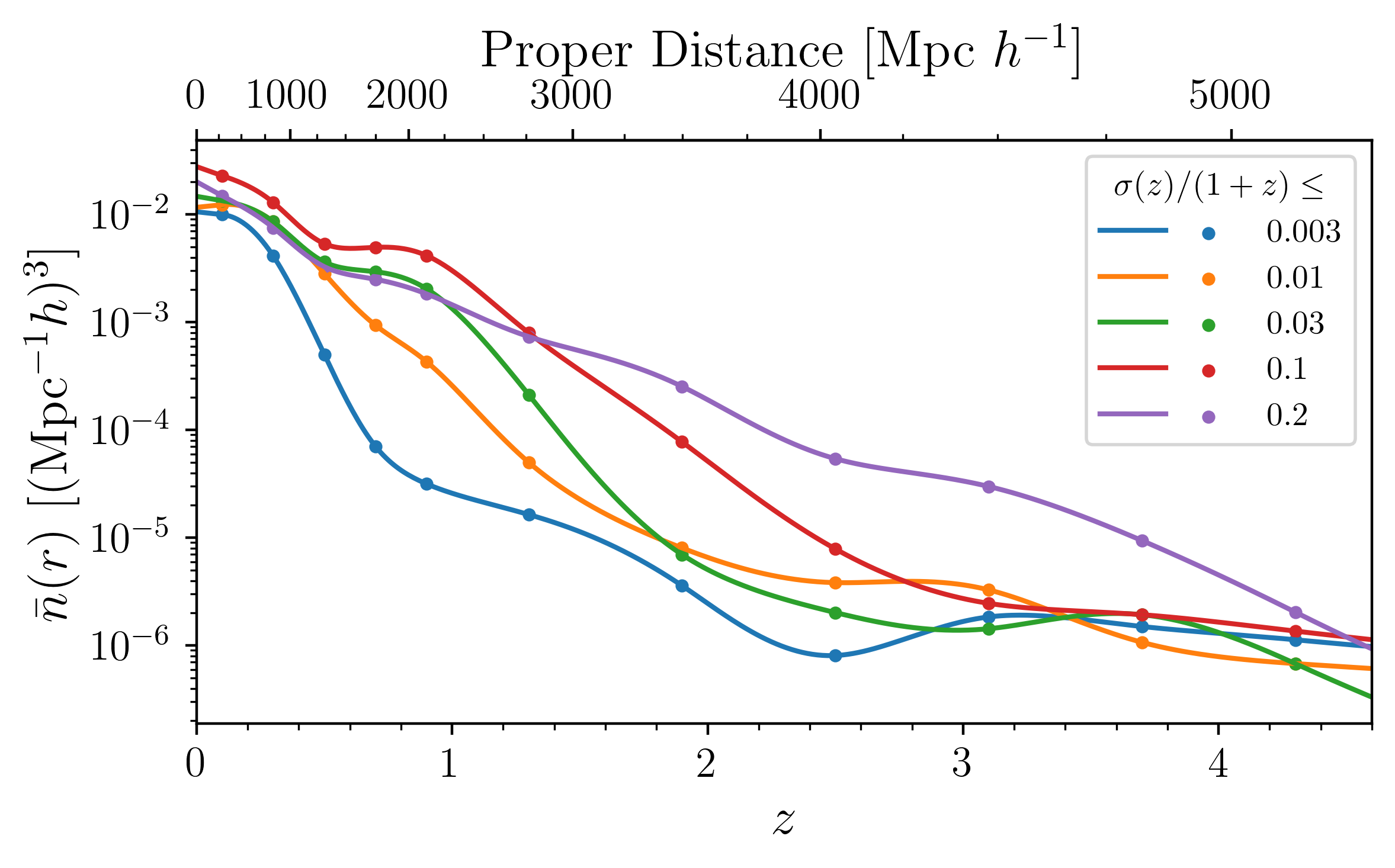}
    \caption{Average galaxy number density data used throughout this paper. A cubic spline was fit to the five SPHEREx sample number density estimates shown as points in the plot.
    }
    \label{fig:sel_func}
\end{figure}
\begin{figure*}[!htbp]
    \centering
    \includegraphics[width = .6\textwidth]{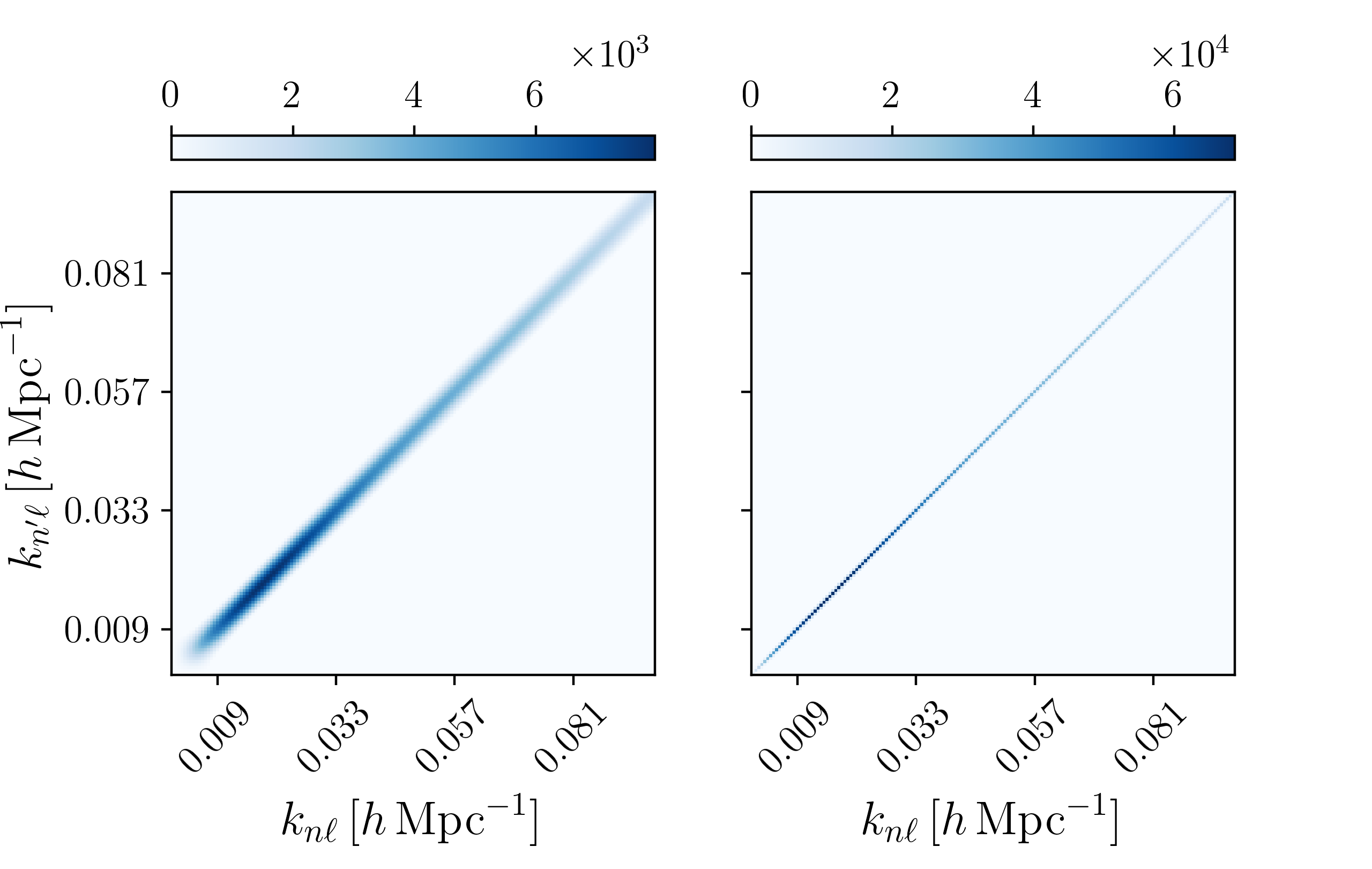}
    \caption{SFB power spectrum $C_\ell(\knl, \knlp)$ in $\left[h^{-1} \text{Mpc}\right]^3$ for $\ell=2$ with Left: SPHEREx-based splined radial selection function. Right: Constant selection function $\phi(r) = 1$.
    }
    \label{fig:2d-cl-sel-func}
\end{figure*}

The observed galaxy number density $\bar n$ is not spatially homogeneous due to inevitable magnitude limitations in galaxy surveys, resulting in more nondetections at higher redshifts. This bias toward galaxies proximal to the survey's minimum radius can be expressed in the form of the radial selection function $\phi$. A plot of the estimated SPHEREx average galaxy number density, separated into five subsamples by forecasted redshift uncertainty $\sigma(z)/(1+z)$, is given in \cref{fig:sel_func}. The curves are obtained by splining the data \cite{spherexpublicproducts} in log-space and are extrapolated linearly. We define the selection function as $\phi(r) = \bar n(r)/\bar n_{\text{0}}$ where the normalization $\bar n_0$ is the maximum value attained by $\bar n(r)$ in the survey range.

In real space, a homogeneous and isotropic matter overdensity field results in a highly localized Fourier transform with power only along the diagonal $\knl = \knlp$. In this case, the SFB modes, and therefore clustering at scales $\knl$ and $\knlp$, are uncorrelated. However, inhomogeneity on the light cone breaks this symmetry. In \cref{fig:2d-cl-sel-func}, we compare the effects of a constant and SPHEREx-based selection function on the SFB power spectrum. The latter results in more power for off-diagonal terms $\knl \neq \knlp$ than the former due to the observed radial inhomogeneity as the observed galaxy number density decreases at farther distances. The correlation between SFB modes is a result of the broken translational invariance in the real-space overdensity field. In other words, following the uncertainty principle, the lack of homogeneity in real space no longer results in a SFB transform with power at a single $\knl$ frequency, but instead we obtain coupled SFB coefficients.

\subsection{Redshift-space distortions}
\begin{figure*}[!htbp]
    \centering
    \includegraphics[width=.75 \textwidth]{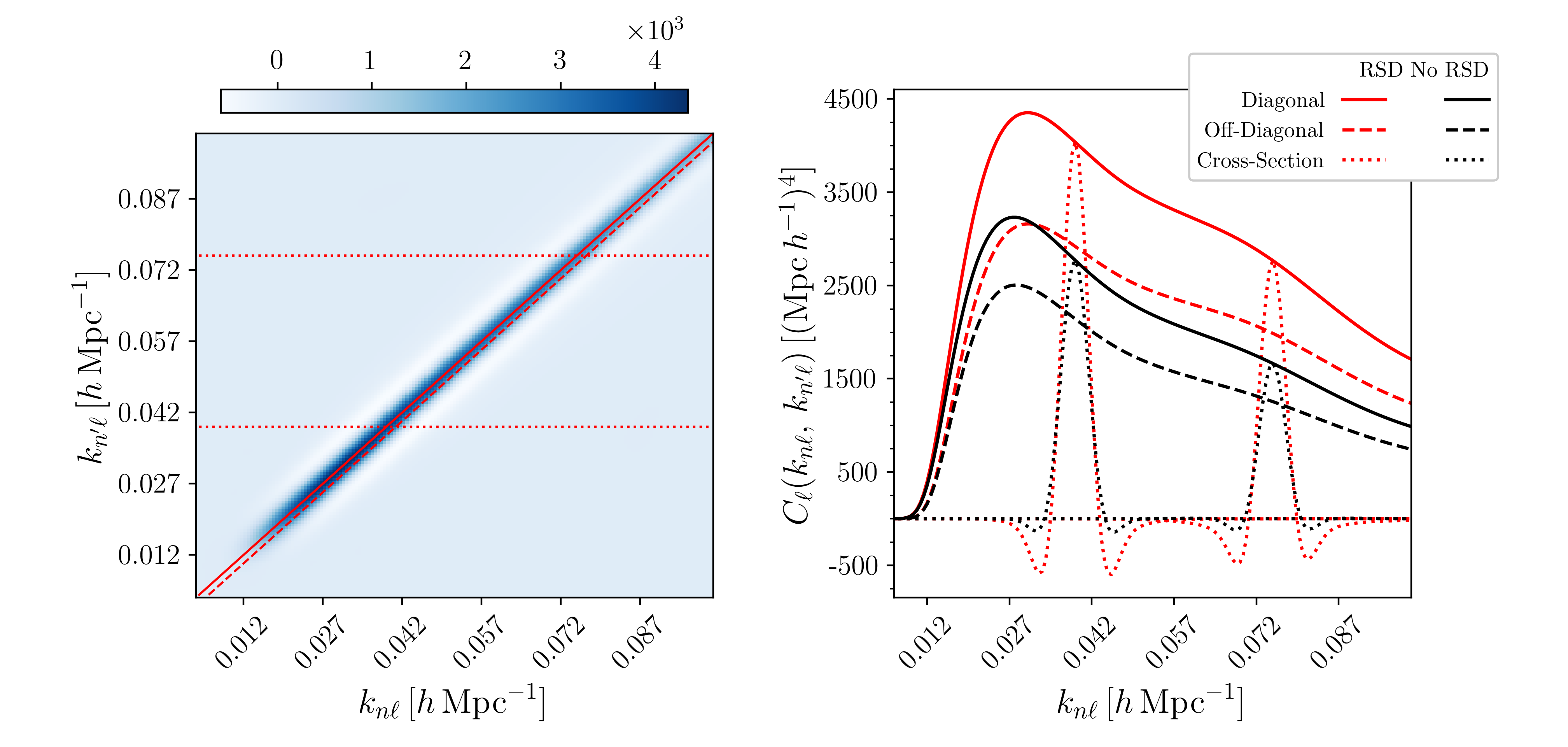}
    \caption{SFB power spectrum for $\ell =10$ with and without RSD. Left: Heatmap of the SFB power spectrum with RSD included. One-dimensional cuts are highlighted in red and displayed in the right plot. The same cuts are taken from the SFB power spectrum without RSD. Right: One-dimensional cuts of the SFB power spectrum. Red lines are slices of the SFB power spectrum in redshift space, and black is without RSD. Solid lines are diagonal cuts, dashed lines indicate off-diagonal cuts, and dotted lines are cross-sections.}
    \label{fig:2Drsd}
\end{figure*}
\begin{figure}
    \centering
    \includegraphics[width=.45 \textwidth]{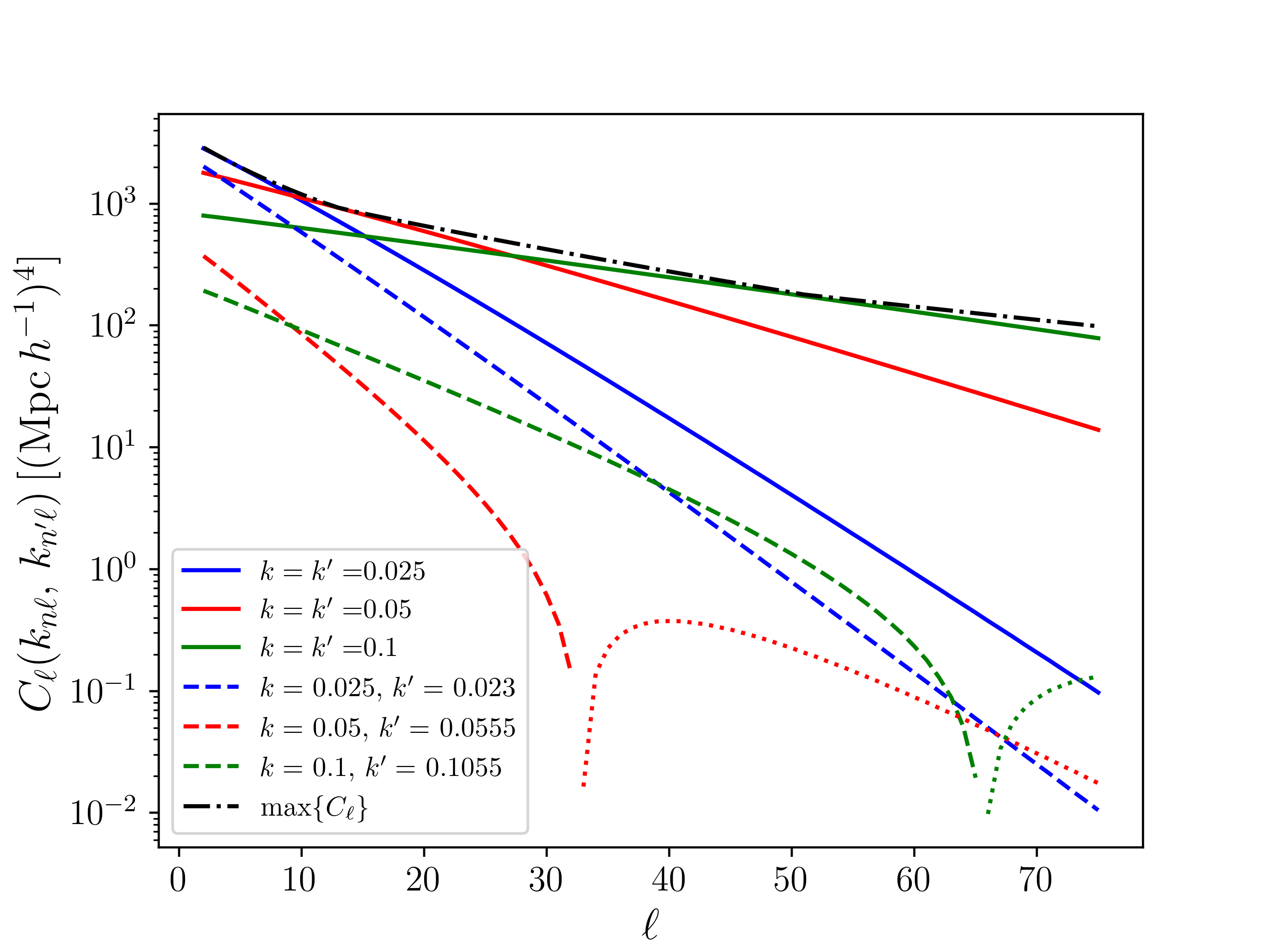}
    \caption{The SFB power spectrum at a fixed $k$ and $k'$ with varying $\ell$. Lines of constant slope appear due to the dominating decay rate of the exponential selection function. The power spectrum becomes negative on the off-diagonal, so the absolute value of the amplitude is displayed as the dotted lines.}
    \label{fig:ellevolution}
\end{figure}

Peculiar velocities bias estimates of distances to galaxies when using redshifts. In particular, in large-scale structures, the infall of galaxies toward a common gravitational center results in a redshift of light emitted from galaxies infalling along the LOS direction and blueshift of those antiparallel to it. This LOS effect, known as the Kaiser effect, results in inferences of more clustering on large scales than is present in real space. A consequence of RSD is therefore an increase in the clustering amplitude of the power spectrum on large scales, as shown in \cref{fig:2Drsd}. While the power is increased for terms adjacent and parallel to the diagonal, the width of the diagonal itself remains relatively unchanged. However, a signature of RSD is the anticorrelation (negative power) in off-diagonal terms near the diagonal. The slight shift of the SFB power spectrum amplitude to larger $\knl$ with RSD can be attributed to the ``squashing'' of large-scale structures to smaller scales from the LOS component of the galaxies' peculiar velocities.

We can also observe where the power resides as a function of $\ell$ for fixed $k$ and $k'$. This can already be seen comparing \cref{fig:2d-cl-sel-func} and \cref{fig:2Drsd} which correspond to $\ell=2$ and $\ell=10$, respectively; the maxima of the power spectra shift to smaller scales (larger $\knl$) as $\ell$ increases. To illustrate this further, in \cref{fig:ellevolution}, the SFB power spectrum is shown for $\ell=2,\dots,75$ in redshift space with an exponential selection function $\phi(r) = \exp(-0.0019 r)$. Small $\ell$ probes larger angular scales, resulting in more power at small $k$ and $k'$ which decreases rapidly as $\ell$ increases. Conversely, larger $\ell$, or smaller angular scales, finds more power in smaller radial scales, so we find that the peak of the power spectrum shifts to larger $k$ as $\ell$ increases. The lines of constant slope can be explained by applying Limber's approxmiation \citep{limber,LoVerde+:2008PhRvD..78l3506L}
\begin{align}
    j_\ell(kr) \approx \sqrt{\frac{\pi}{2kr}}\delta^D\left(kr-\ell-\frac{1}{2}\right). \label{eq:limber}
\end{align}
to \cref{eq:wnl,eq:cl}. The exponential selection function dominates the decay rate, resulting in straight lines. The magnitude of the slopes for different $k$ is also explained by the argument to the selection function in this approximation. 

\subsection{Non-Gaussianity}
\begin{figure*}
    \centering
    \includegraphics[width=.75\textwidth]{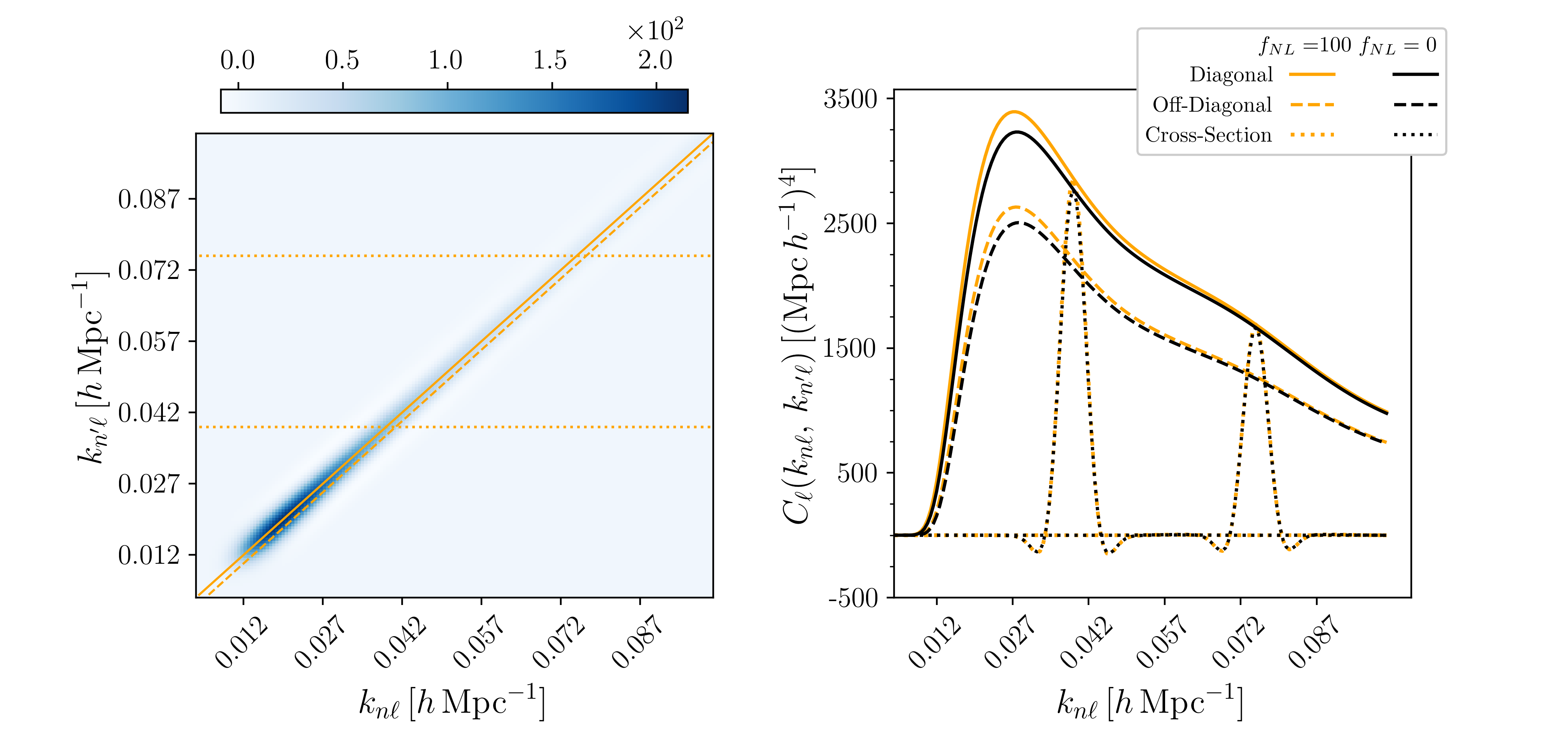}
    \caption{SFB power spectrum in real space for $\ell =10$ with and without non-Gausianity. Left: Difference between the SFB power spectrum with $\fnl=100$ and with $\fnl=0$. One-dimensional cuts are highlighted in gold and displayed in the right graph with the undifferenced power spectra.
    Right: One-dimensional cuts of the SFB power spectrum. Gold lines are slices with $\fnl=100$, and black is with $\fnl=0$.}
    \label{fig:2Dfnl}
\end{figure*}
The slow-roll, single-field inflationary paradigm predicts a mostly Gaussian random density field \cite{planckInflation}. Constraining the level of non-Gaussianity (NG) of matter overdensities serves in delineating which inflationary models best describe the complex physical processes of the early universe. In some of these models, NG can be captured by the SFB power spectrum as an observable on large scales. This is due to the fact that large-scale structure falls on the tail of the probability distribution of the real space density contrast, making it especially sensitive to the skewness resulting from local NG \cite{Dalal:2007cu}.

We adopt a galaxy bias with a scale-dependent correction term \citep{Zhang:2021wzo}
\begin{align}
    \Delta b(r, k) = 2(b-1) f_{\text{NL}}\delta_c\frac{3\Omega_m}{2\widetilde D(r) T(k) r_H^2 k^2}\,,
    \label{eq:bias_correc}
\end{align}
to measure the SFB power spectrum under the assumption of local primordial non-Gaussianity. Here, $b$ is the fiducial galaxy bias, $T(k)$ is the matter transfer function, $\widetilde D(r)$ is the growth factor normalized to the scale factor during matter domination, $r_H$ is the Hubble radius, $\fnl$ characterizes the dependence of the amplitude on non-Gaussian effects, and we take the critical overdensity as $\delta_c = 1.686$.

\cref{fig:2Dfnl} displays how NG is realized in the SFB PS. In the plot, $\fnl=100$ is exaggerated from typically constrained values \cite{planckfnl} for demonstration. The power spectrum amplitude is increased at very large scales $\knl \lesssim 0.05 h$ Mpc$^{-1}$. This scaling is maximized along the diagonal, and off-diagonal terms do not see as large of an increase in power. The width of the diagonal shows no discernible change with nonzero $\fnl$. Most of the information to be extracted from the SFB power spectrum with regard to NG lies at small $k=k'$ and $\ell$ values.

\subsection{Shot Noise \label{sec:shotnoise}}
\begin{figure*}
    \centering
    \includegraphics[width=\textwidth]{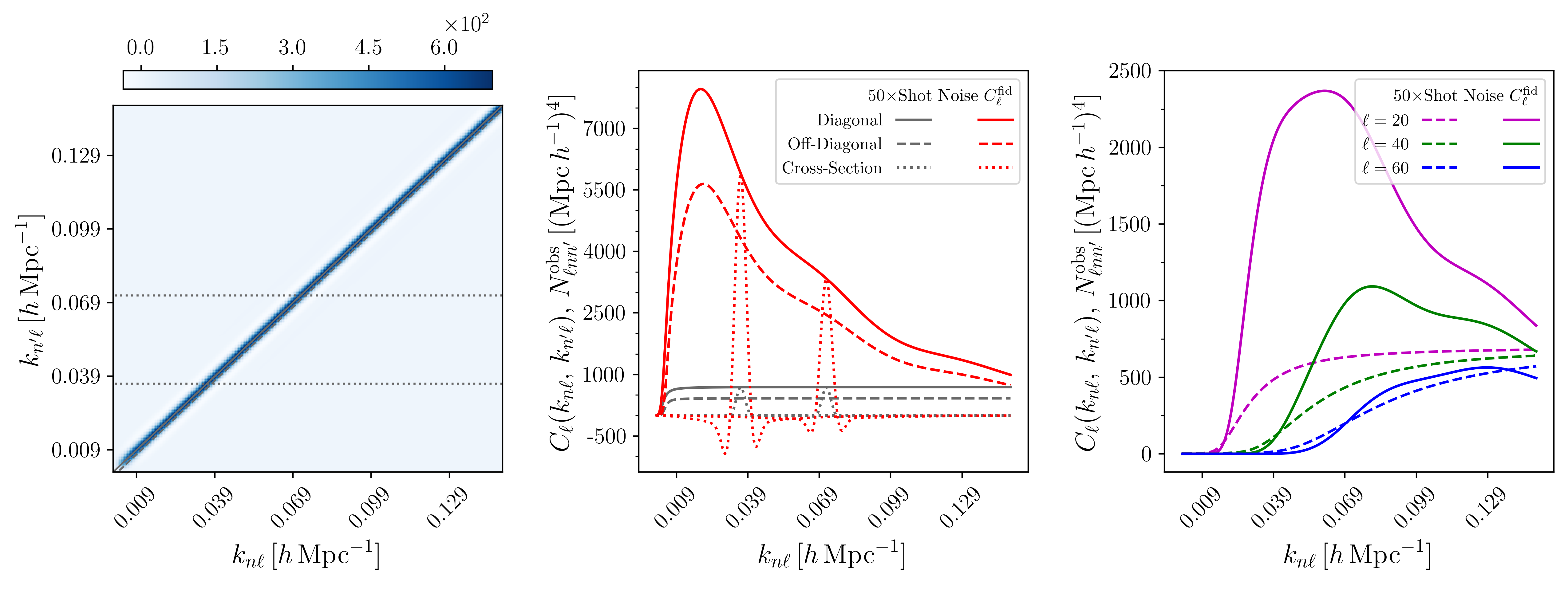}
    \caption{Comparison of the shot noise, multiplied by a factor of 50 in each plot, and the SFB power spectrum for $\ell =2$. Left: Shot noise matrix. One-dimensional cuts are shown in gray and displayed in the middle graph. Middle: One-dimensional cuts of the shot noise and SFB power spectrum with RSD. Gray lines are slices from the left plot, and red are the same cuts from the SFB PS. Right: Rescaled shot noise (dashed) and SFB power spectrum (solid) for larger $\ell$ along the diagonal.}
    \label{fig:shotnoise}
\end{figure*}
Studying the overdensity field in the cosmic variance-limited case is instructive for isolating and exploring the manifestations of physical effects in the SFB power spectrum. However, in order to perform meaningful statistical analyses, we must account for shot noise, which increases the uncertainty in measurements of quantities such as the cosmological parameters due to the sampling of finitely many galaxies within the survey volume. From \cite{superfab}, the shot noise matrix elements are given via 
\begin{align}
    N_{\ell nn'}^\mathrm{obs}
    &=
    \frac{1}{n_0}\frac{1}{\sqrt{4\pi}}
    \int dr\,r^2\,g_{n\ell}(r)\,g_{n'\ell}(r)
    \,W_{00}(r)\,,
\end{align}
where, assuming that the full window $W_{00}$ is separable into a mask $M(\hat{\mathbf{r}})$ and radial selection $\phi(r)$,
\begin{align}
    W_{00}(r)
    &= \int d\Omega \,Y^*_{00}(\theta,\varphi)\,M(\theta,\varphi)\,\phi(r)
    \\
    &\simeq \sqrt{4\pi}\,\phi(r)\,f_\mathrm{sky}\,,
\end{align}
for fractional sky coverage $f_\mathrm{sky}$. Thus,
\begin{align}
    N_{\ell nn'}^\mathrm{obs}
    &=
    \frac{f_\mathrm{sky}}{n_0}
    \int dr\,r^2\,g_{n\ell}(r)\,g_{n'\ell}(r)
    \,\phi(r), \label{eq:shotnoise}
\end{align}
where $n_0$ is the maximum galaxy number density within the survey. The shot noise matrix is then simply added to the power spectra. In this work, the window function consists only of the radial selection function, and we approximate the fractional sky coverage after calculating the Fisher information as shown in the next section, thereby setting $\fsky=1$ in \cref{eq:shotnoise} to avoid accounting for the angular selection function twice. We leave a more precise treatment of the angular mask for future work.

For demonstration, $N^\mathrm{obs}_{\ell n n'}$ for $\ell=2$ is displayed in \cref{fig:shotnoise}. The shot noise dominates at smaller scales where it asymptotes to a constant. This can be explained by the fact that in \cref{eq:shotnoise}, the $g_{n\ell}$ remain approximately zero until peaking around $r_\mathrm{peak}=(\ell+1/2)/k_{n\ell}$, which is found using the Limber approximation \cref{eq:limber} in the case that $r_\mathrm{min} \approx 0$. When $k_{n\ell}\sim \mathcal{O}(10^{-3})h/\mathrm{Mpc}$, the peak lies outside the survey range since $r_\mathrm{peak}\gg r_\mathrm{max}$. Thus, increasing $k_{n\ell}$ shifts the peak to smaller $r$ so that oscillations begin to enter the survey boundary and contribute to the integral. It is easier to see why this contribution converges at large $k_{n\ell}$ when $\phi=\mathrm{const}$ and $r_\mathrm{min}=0$ so that the radial eigenfunctions are the spherical Bessel functions $j_\ell$. At large $k_{n\ell}\,r$, $j_\ell(k_{n\ell}\,r)$ scales as $1/r$, so the integrand is approximately the square of sinusoidal function. Averaging this function from $\rmin$ to $\rmax$ then gives a constant approximately independent of $k_{n\ell}$ because the integrand's wavelength is much smaller than the survey's radial length.

\section{Fisher Matrix Forecasting\label{sec:fisher}}

We will now leverage our code to predict the considered survey constraints on cosmological parameters $\bm{\theta}=( b\sigma_8, f\sigma_8, h, \Omega_m, \fnl)$. We review the Fisher forecasting formalism before presenting our results. 

We construct the likelihood function by defining the data vector in SFB space, $\boldsymbol{\hat\delta}\equiv \delta_{n\ell m}^\mathrm{obs}$. Then,
the probability of measuring this data vector given parameters $\boldsymbol{\theta}$ is given by the Gaussian likelihood function
\begin{align}
    \label{eq:likelihood}
    \boldsymbol{\mathcal{L}}(\boldsymbol{\hat \delta} \,|\,\boldsymbol{\theta}) &=\frac{1}{(2\pi)^{\dim (\boldsymbol{\hat \delta})/2}\sqrt{|\boldsymbol{\mathcal{C}}|}} \exp\left\{-\frac{1}{2}\boldsymbol{\hat \delta}\,\boldsymbol{\mathcal{C}}^{-1}\boldsymbol{\hat \delta}\right\},
\end{align}
where $\boldsymbol{\mathcal{C}} \equiv C_{\ell}(\knl,\knlp) + \nobs$ is the discretized SFB power spectrum with shot noise \cref{eq:shotnoise}. We denote $\theta_n$ as the $n^{\text{th}}$ parameter in $\boldsymbol{\theta}$ and define $\square_{,\alpha} \equiv \partial\square/\partial\theta_\alpha$. Then, the Fisher matrix element at indices $\alpha$ and $\beta$ is given by
\begin{align}
    F_{\alpha\beta} =  \left\langle -(\ln\boldsymbol{\mathcal{L}})_{,\alpha\beta} \right\rangle.
\end{align}
which, through \cref{eq:likelihood}, can be rewritten as
\begin{align}
    \label{eq:fisherinformation}
    F_{\alpha\beta} = f_{\mathrm{sky}}\sum_{\ell = \ell_{\text{min}}}^{\ell_{\text{max}}}\frac{2\ell+1}{2}\text{tr}\left(\boldsymbol{\mathcal{C}}^{-1}\boldsymbol{\mathcal{C}}_{,\alpha}\boldsymbol{\mathcal{C}}^{-1}\boldsymbol{\mathcal{C}}_{,\beta}\right),
\end{align}
given a fractional sky coverage $f_{\mathrm{sky}}$ and $2\ell +1$ $m$-modes per $\ell$. Here, we choose $\ell_\mathrm{min}=2$ to avoid systematics associated with the monopole and dipole contributions, such as our own peculiar velocity. Additionally, we choose $\ell_\mathrm{max}$ \textit{a posteriori} with observation of when the $\ell\approx \ell_\mathrm{max}$ mode contributions to the reduction of errors on $\boldsymbol{\theta}$ become negligible, as will be later shown in \cref{fig:uncertainty_evol}. Finally, the covariance matrix of the parameters $\boldsymbol{\theta}$ is obtained via the inversion of the Fisher information matrix
\begin{align}
    C_{\alpha\beta} =(F^{-1})_{\alpha\beta}.
\end{align}
Derivatives of the power spectra are computed numerically by inducing small perturbations in the cosmological parameters $\alpha,\beta \in \boldsymbol{\theta}$, then dividing the differenced fiducial and perturbed power spectra by the perturbation size. This is straightforward for $h$, $\Omega_m$, and $\fnl$, though for $\fs(z)$ and $\bs(z)$, we model deviations from the reference cosmology by a low-order polynomial in redshift as described in \cref{sec:zevol}.

In order to compute the numerical derivatives of the power spectrum with respect to parameters $\Omega_m$ and $h$, we must calculate the input matter power spectrum in a universe different than the \textit{Planck} flat $\Lambda$CDM model used throughout this work. The perturbed parameters result in a different set of basis functions for the SFB decomposition and therefore different wavenumbers $k$. Thus, to difference SFB power spectra between cosmologies, we calculate the SFB power spectrum in the perturbed cosmological model, perform a two-dimensional spline, and evaluate the splined SFB power spectrum at the allowed wavenumbers in the fiducial cosmology.

\subsection{Modeling the Redshift Evolution \label{sec:zevol}}
Here, we describe the model used for $\fs$ and $\bs$ in our Fisher matrix calculation since these are functions of redshift. We only explicitly write the model for $\fs$ because $\bs$ has the same form, so the substitution $\fs\rightarrow\bs$ and $f(z)\rightarrow b(z)$ can be made to obtain the equations for $\bs$.

Given a fiducial form $\fs^\mathrm{fid}(z)=f(z)\,D(z)\,\sigma_8$ where $\sigma_8$ is evaluated at redshift $z=0$, we model the true functional form by a modulation with a low-order polynomial. That is,
\begin{align}
    f\sigma_{8}(z) \equiv \fs^\mathrm{fid}(z) \sum_{i=0}^n a_i z^{i}\,, \label{eq:fbsig8_mod}
\end{align}
where $a_i$ are the parameters to be fit and $n$ is the order of the polynomial. The goal is then to obtain uncertainties for the parameters $a_i$ from our Fisher analysis.

To obtain an uncertainty forecast for $\fs(z)$ itself, we propagate the uncertainties in the $a_i$ to linear order. That is, we assume that small changes $\Delta a_i$ will lead to a change $\Delta\fs(z)$
\begin{align}
    \Delta\fs(z)
    \approx
    \sum_{i=0}^n \frac{\partial \fs(z)}{\partial a_i} \, \Delta a_i\,.
    \label{eq:fbsig8_approx}
\end{align}
The variance in $\fs(z)$ is then given by
\begin{align}
    \langle(\Delta\fs)^2\rangle
    &=
     \left[\fs^\mathrm{fid}(z)\right]^2 \sum_{ij} z^{i+j} \left\langle\Delta a_i\,\Delta a_j\right\rangle,
    \label{eq:fsig8_variance}
\end{align}
where the sum is over $i,j=0,\ldots,n$ and we have assumed the modulated model \cref{eq:fbsig8_mod}.
The covariance between the coefficients $a_i$ and $a_j$ is found from the inversion of the Fisher information matrix
\begin{align}
    \langle \Delta a_i\,\Delta a_j \rangle
    =
    \left(F^{-1}\right)_{ij}.
\end{align}
Therefore, by evaluating the Fisher information matrix, we can find the uncertainty in $\fs$ as a function of redshift. However, we stress that in a measurement the parameters $a_i$ are constrained rather than $f\sigma_8(z)$ at some discrete redshifts, and the constraints on the parameters $a_i$ come from all redshifts.

\subsection{\label{sec:cov}Fisher Forecast Results}
\subsubsection{Single Galaxy Subsample}
\begin{figure}
    \centering
    \includegraphics[width = .5 \textwidth]{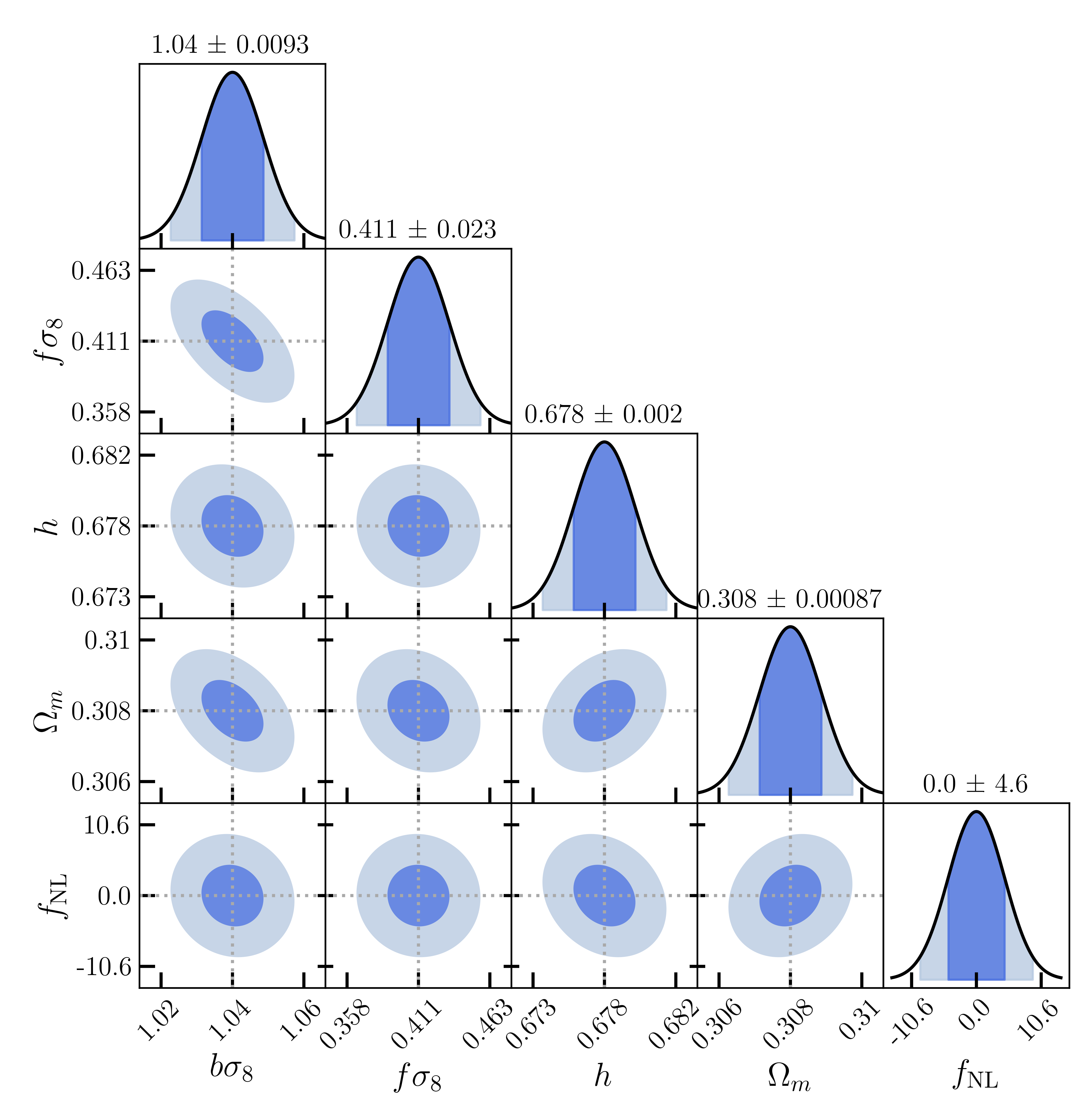}
    \caption{Forecasted constraints on cosmological parameters at $z=0$ in a survey from $z = 0$ to 4.6 with $k_{\text{max}}= 0.15\,h/\text{Mpc}$ and $\ell = 2,\,\dots,\,300$ and 80\% sky coverage and redshift errors $\sigma_z/(1+z)\leq 0.003$. Dark blue and light blue indicate the 68\% and 95\% confidence level regions, respectively. For $\fs$ and $\bs$, we expand to fifth order in $z$ in \cref{eq:fbsig8_mod}. Other parameters such as the spectral tile $n_s$ are fixed and not marginalized over.}
    \label{fig:corner}
\end{figure}
\begin{figure*}
    \centering
    \includegraphics[width = .48 \textwidth]{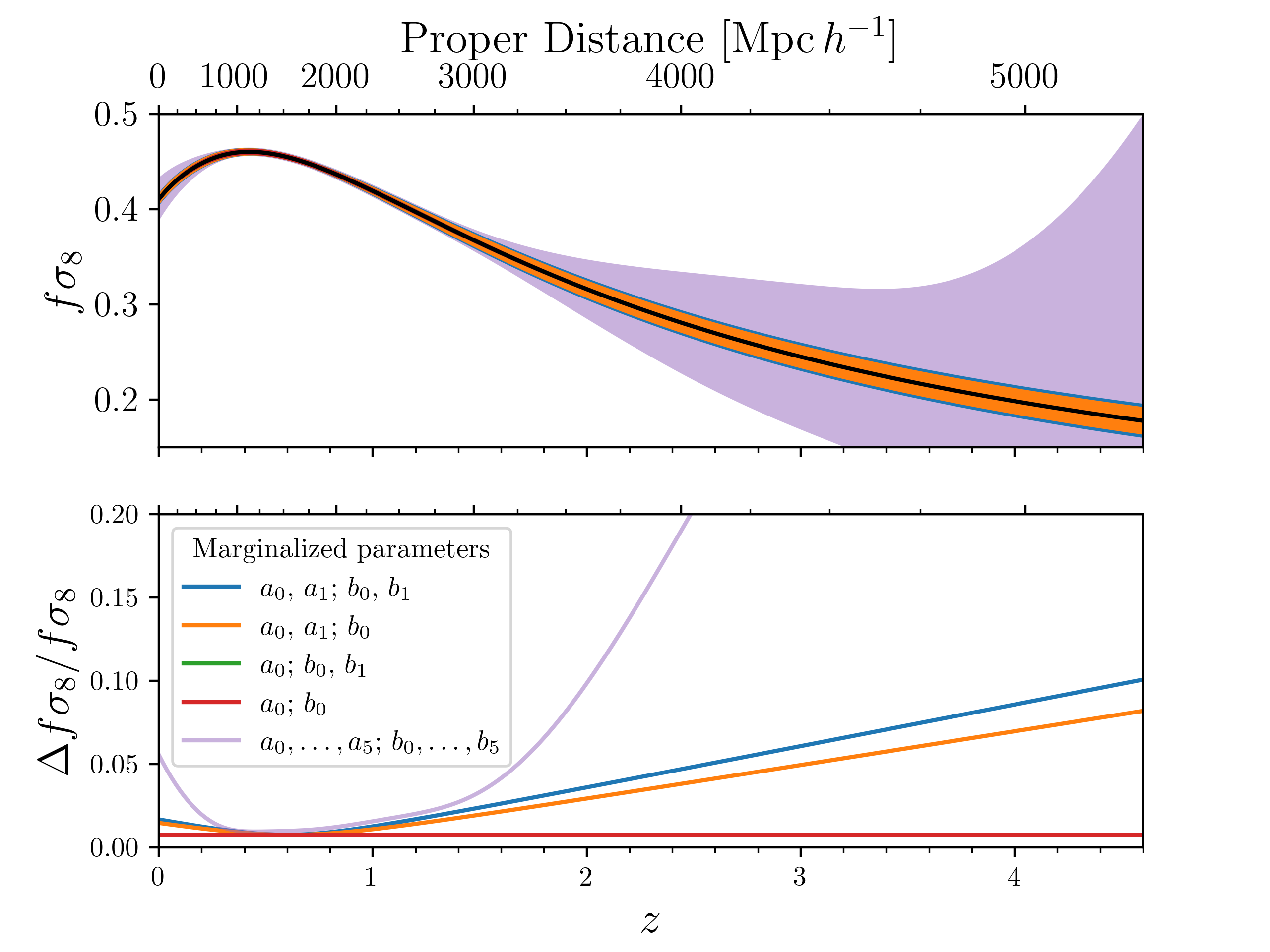}
    \includegraphics[width = .48 \textwidth]{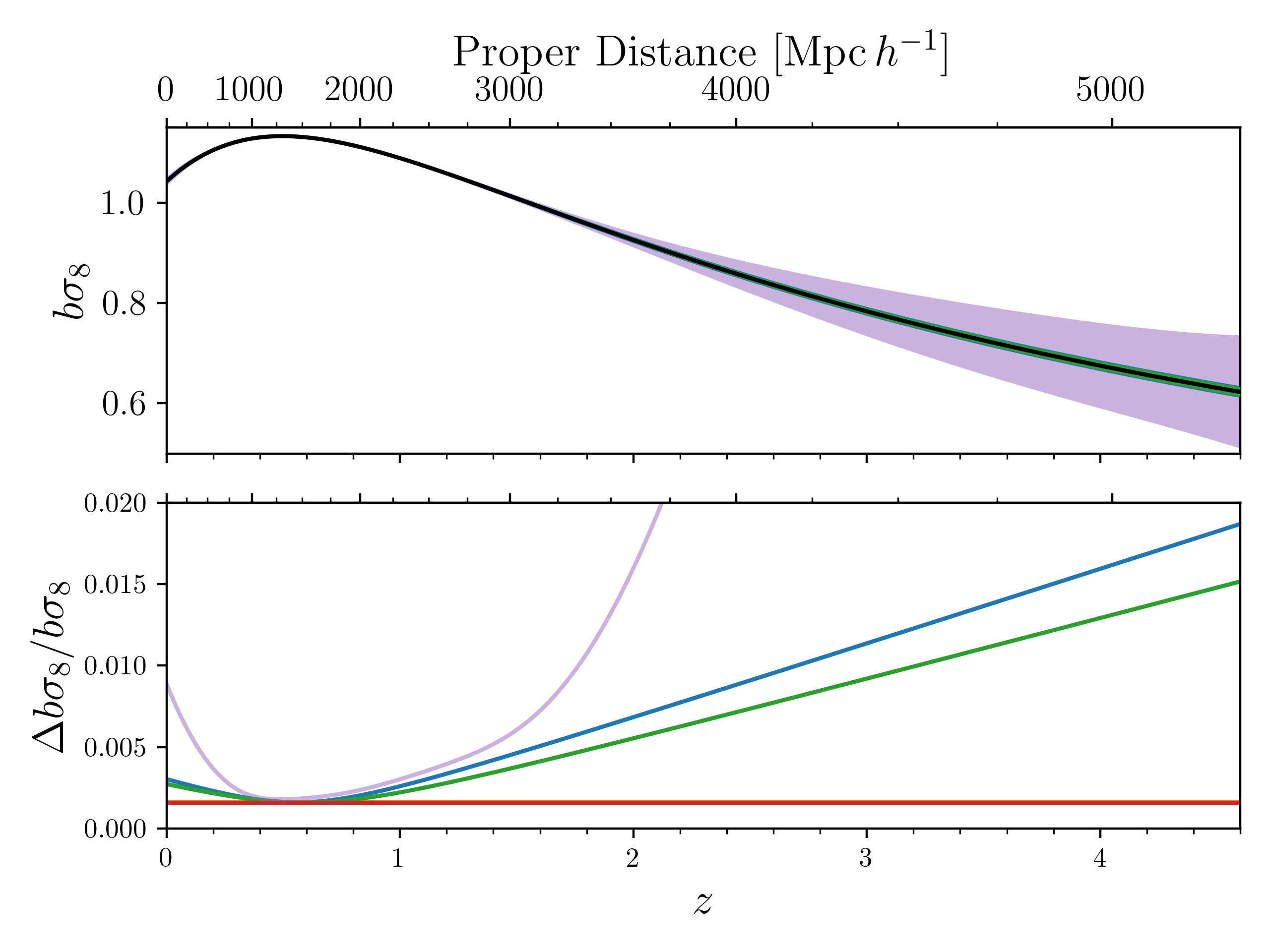}
    \caption{Redshift evolution of $f\sigma_8$ and $\bs$ from the Fisher forecast for $\ell=2,\dots,300$ for redshift errors $\sigma_z/(1+z)\leq 0.003$. The 68\% confidence level band is shown as the shaded region in top panel of both plots. The normalized error is plotted alone in the bottom panels. The legend refers to coefficients in \cref{eq:fbsig8_approx} for $\fs$ as $a_i$ and as $b_i$ for $\bs$. Left: The error curve for $f\sigma_8(z)$. The green error curve is approximately the same as the red curve. Right: The same plot as the left, though for constraints on $\bs$(z). The error curve in orange is approximately the same as that in red.}
    \label{fig:fbsig8}
\end{figure*}
\begin{figure}
    \centering
    \includegraphics[width = .5 \textwidth]{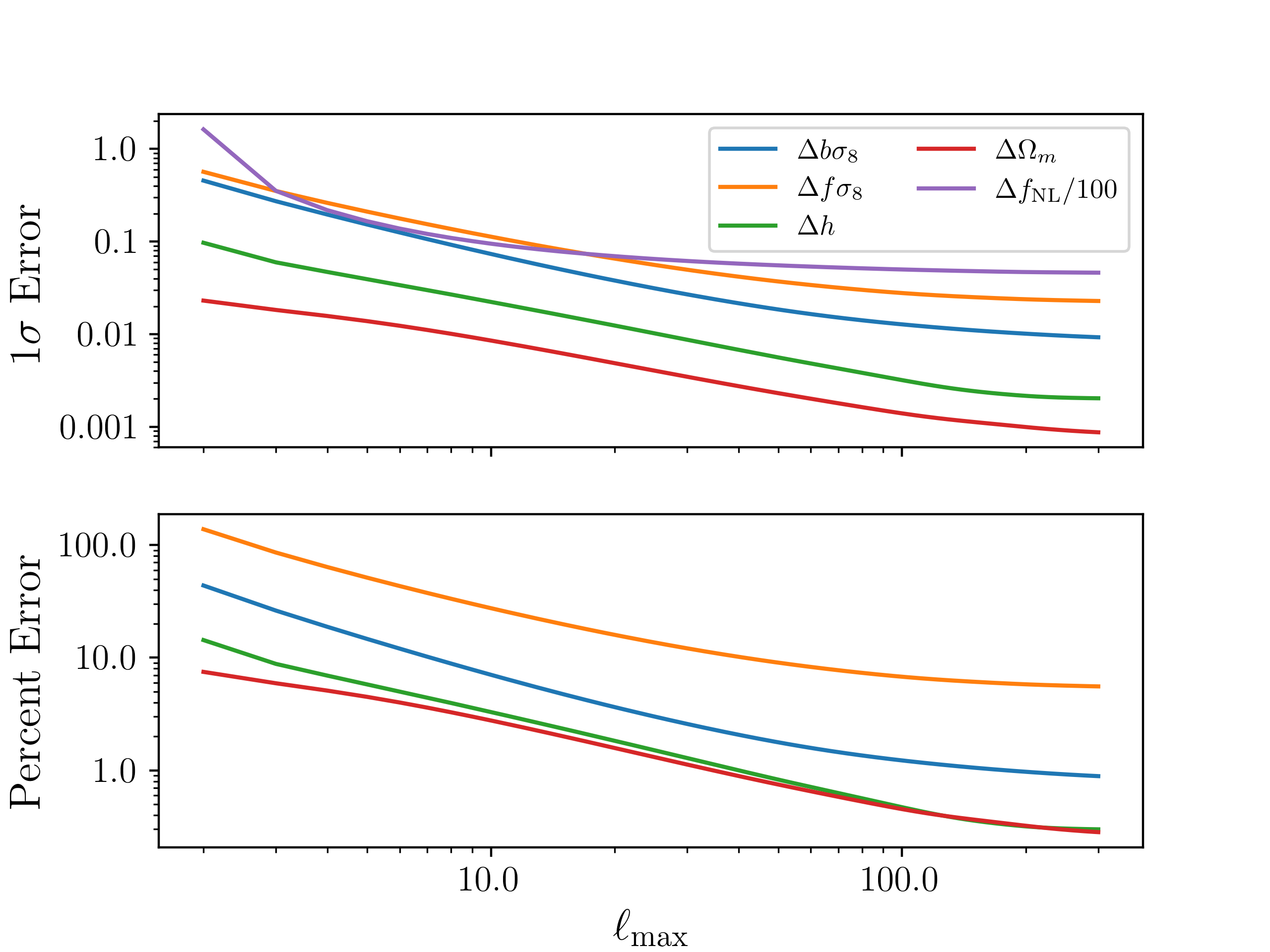}
    \caption{Constraints on cosmological parameters as a function of the maximum $\ell$ mode used in the forecast, starting from $\ell=2$ and perturbing $\fs,\,\bs$ to 5th order in $z$. Note that the constraint on $\fnl$ has been scaled to fit the top plot.}
    \label{fig:uncertainty_evol}
\end{figure}
We present forecasts for the survey we consider in \cref{fig:corner} where we have adopted the galaxy bias correction due to local primordial non-Gaussianity given in \cref{eq:bias_correc} for our estimates of $\fnl$ having fixed the spectral tilt $n_s=0.965$ and $\sigma_8^{(0)}=0.7886$. We have forecasted $b\sigma_8 = 1.042\pm 0.0093$, $f\sigma_8=0.4106\pm 0.023$, $h=0.6778\pm0.0020$, $\Omega_m = 0.3081\pm 0.00087$, and $\fnl=0\pm 4.6$. The uncertainties for $\bs$ and $\fs$ are obtained by modulating the perturbative expansion of these models to fifth order. Furthermore, we take $\fsky = 0.8$ and use a selection function splined in log-space from the public data described in \cref{sec:sel_functn}. The bias model is obtained by performing linear regression on the data \cite{spherexpublicproducts} as well. The constraints on $\Omega_m$ and $h$ are likely tight due to the fact that we fix the spectral tilt $n_s$ to the fiducial value.

\cref{fig:fbsig8} shows the evolution of $f\sigma_8$ and $b\sigma_8$ within the survey range with a $1\sigma$ error band from the Fisher forecast. Note that some regions are almost identical and overlap, as described in the caption. At $z \gtrsim 1$, the variance \cref{eq:fsig8_variance} scales quadratically at high $z$ when including terms up to linear order in our polynomial expansion, thus, the standard deviation evolves approximately linearly at large $z$. The minimum of the errors at $z\approx 0.5$ can be explained by the fact that the covariance $\langle a_0 a_1\rangle $ is less than zero, so we have a negative term at linear order in $z$ in \cref{eq:fbsig8_approx} which shifts the minimum of $\sqrt{\langle (\Delta \fs)^2\rangle}$ to the right; the same reasoning follows for $\bs$.

\cref{fig:uncertainty_evol} indicates that most of the information from our analysis has been extracted from the power spectrum. Multipoles above $\ell=200$ contribute negligibly to reducing error estimates. In particular, uncertainty for $\fnl$ is primarily reduced at large angular scales as a result of the scale-dependent bias correction $\Delta b(r, k) \propto k^{-2}$. The imprint of non-Gaussianity on the cosmological overdensity field will primarily manifest itself in the linear regime. 

\subsubsection{Multi-tracer Analysis}
\begin{figure*}
    \centering
    \includegraphics[width=\textwidth]{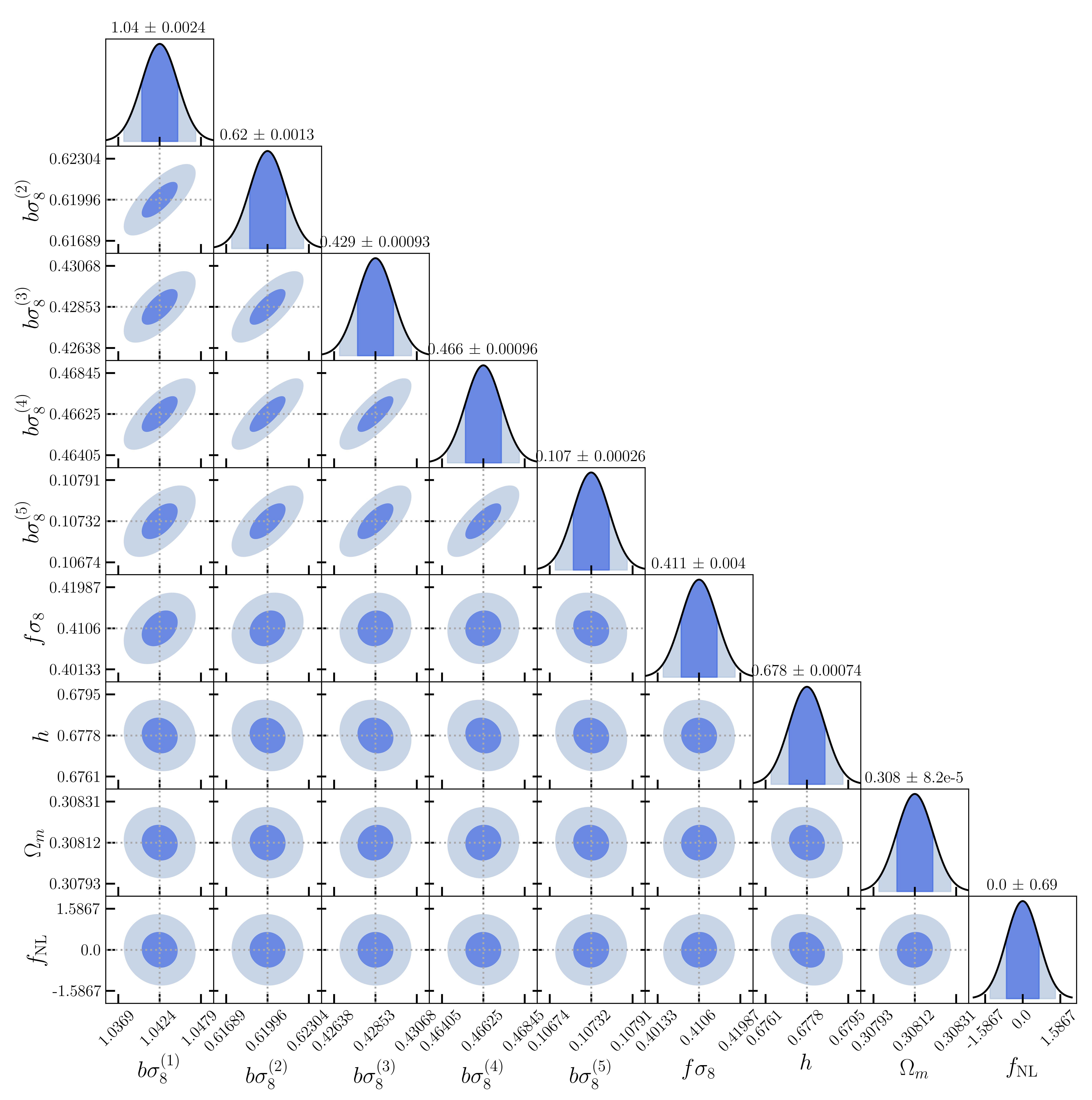}
    \caption{Multi-tracer Fisher forecast at $z=0$. The superscripts in $b\sigma_8^{(i)}$ label which redshift subsample from \cite{spherexpublicproducts} that the bias model corresponds to.}
    \label{fig:multitracer-forecast}
\end{figure*}
\begin{table*}[ht]
    \centering
        \begin{tabular}{@{\extracolsep{4pt}}llccccccc}
        \toprule   
        {} & \multicolumn{5}{c}{Subsample Uncertainty Upper Bound $\sigma(z)/(1+z)$} & {}\\
         \cmidrule{2-6}
         Parameter & $ 0.003$ &  $ 0.01$ &  $ 0.03$ &  $ 0.1$ &  $ 0.2$ & {Cross-Correlation} \\ 
        \midrule
        $\Delta b\sigma_8\times 10^3$& 9.257 &6.515 & 7.272 & 5.794 & 4.805 & See \cref{fig:multitracer-forecast} \\ 
          $\Delta f\sigma_8\times 10^3$& 22.82 & 15.66 & 13.6 & 11.79 & 7.921 & 4.030 \\ 
          $\Delta h\times 10^3$  & 2.03 & 1.189 & 1.007 & 0.8576 & 0.7961 & 0.7382 \\ 
          $\Delta\Omega_m\times 10^3$  & 0.8715 & 0.5112 & 0.4326 & 0.2636 & 0.1135 & 0.08197 \\ 
          $\Delta f_\mathrm{NL}$  & 4.614 & 2.202 & 2.055 & 1.381 & 0.779 & 0.6899 \\ 
        \bottomrule
        \end{tabular}
    \caption{$1\sigma$ forecasted errors ($\times 10^3$, except for $f_\mathrm{NL}$) on cosmological parameters from the Fisher forecasts at $z=0$ marginalizing over all $b_i$ for $\ell = 2,\dots,300$. $\fs$ and $\bs$ are expanded to 5th order in $z$. The last column indicates constraints using cross-correlations between redshift subsamples.}
    \label{tab:fiveforecast}
\end{table*}
We now present Fisher constraints for a multi-tracer analysis including all five subsamples \cite{spherexpublicproducts} of SPHEREx. Multi-tracer analysis is expected to be a powerful probe of local non-Gaussianity as it reduces the cosmic variance contribution to the measurement error by correlating tracers with different biases \citep{Seljak:2009PhRvL.102b1302S}.

The SPHEREx samples are defined by their redshift uncertainties. Each sample's $b\sigma_8(z)$ is modeled as in \cref{eq:fbsig8_mod} to fifth order. For the cross-correlations, the SFB power spectrum can have two different $\mathcal{W}_{n\ell}(q)$ kernels in \cref{eq:cl} from different subsamples, thus they contain different galaxy biases and selection functions.

To better understand the effect of the multi-tracer analysis, we first perform the Fisher analysis on each SPHEREx galaxy subsample separately, then perform the full analysis including all auto- and cross-correlations between the subsamples. The results are shown in \cref{tab:fiveforecast}, marginalized over the other parameters considered here. A corner plot with these results analogous to \cref{fig:corner} is shown in \cref{fig:multitracer-forecast}.

We find that the multi-tracer analysis significantly improves many of the constraints. In particular, we find that the $\fnl$ constraint is improved from $\pm4.6$ to $\pm0.69$ in the multi-tracer case.

All of these results are evaluated at $z=0$ for a SPHEREx-like survey with 80\% sky coverage and radial extent from $z=0$ to $z=4.6$. We impose high-frequency cutoffs $\ell_\mathrm{max}=300$ and $k_\mathrm{max}=0.15h/\mathrm{Mpc}$. Shot noise terms as in \cref{eq:shotnoise} are included, and high order perturbative expansions of $\fs$ and $\bs$ greater than $\mathcal{O}(z^5)$ are excluded due to numerical instability. We will note that these results are sensitive to the selection function and bias model used, so the accuracy of our forecasts are largely dependent on the behavior of these models. We also limited ourselves to the cosmological parameters shown, and including others will likely degrade some of the forecasted constraints.

\section{\label{sec:conclusions}Conclusions}
We have written code that calculates the SFB power spectrum with discrete SFB modes, and we have leveraged our program to investigate the effects of survey geometry, RSD, and NG on the power spectrum clustering amplitude and the correlation between SFB modes.

The selection function, as well as growth of structure and bias evolution, breaks the homogeneity on our past light cone, resulting in a correlation of terms $\knl\neq\knlp$. Furthermore, the local motion of galaxies within large-scale structure from gravitational clustering results in a LOS component of the galaxies' velocities that contributes to the observed redshift, biasing inferred distance estimates and resulting in an increase of the power spectrum amplitude in redshift space. This also breaks the homogeneity, and results in negative correlation for off-diagonal terms $\knl\neq\knlp$.

Non-Gaussianity, due to the sensitivity of large-scale structure to perturbations in the distribution of real space overdensities, arises on large scales. We also studied the evolution of the power spectrum as a function of multipole moment, and we showed that the typical length scale at which the clustering amplitude is largest is positively correlated with the $\ell$ mode in consideration.

Including shot noise and a spherically symmetric angular mask with 80\% sky coverage, we forecasted uncertainties on the parameters $\fs$, $\bs$, $h$, $\Omega_m$, and $f_\mathrm{NL}$ using the Fisher information matrix formalism. We have also obtained the uncertainties on $\fs$ and $\bs$ as a function of redshift, and by evaluating the contribution of higher multipoles to the constraints, we have verified that information has been maximally extracted from the power spectrum estimator.

We extended our single-tracer Fisher forecasts to perform a multi-tracer forecast with all five SPHEREx subsamples. We showed that the constraints tighten significantly with the cross-correlations.

In future work, we anticipate to include more cosmological parameters in the Fisher forecast, in particular the spectral tilt $n_s$ that is likely responsible for our tight constraints on the $h$ and $\Omega_m$. Since we were limited to a 5th order polynomial for the redshift-evolution of $\fs$ and $\bs^{(i)}$, a possible path forward would be to switch to an expansion in Chebyshev polynomials instead of monomials, which we anticipate will make the numerics more tractable. More general models for the redshift uncertainties will also help make the analysis more realistic.

\begin{acknowledgments}
Part of this work was done at Jet Propulsion Laboratory, California Institute of Technology, under a contract with the National Aeronautics and Space Administration.
\end{acknowledgments}

\appendix

\section{\label{appendix:gaussleg}Gauss-Legendre Quadrature}
The Gauss-Legendre quadrature integration algorithm is crucial for the performance and accuracy of our code. The integral approximation takes the form of 
\begin{align}
    \int_{-1}^1f(x) dx \approx \sum_{i=1}^n w_if(x_i)
    \label{eq:fgq}
\end{align} 
for nodes $x_i$ and weights $w_i$. We utilize the Julia module \texttt{FastGaussQuadrature.jl} \footnote{\url{https://github.com/JuliaApproximation/FastGaussQuadrature.jl}} to calculate the nodes and weights. This approximation is used to compute the integral kernel $\mathcal{W}_{n\ell}$ and the $C_\ell$. The integral defining the kernel has integration bounds $r_{\text{min}}$ and $r_{\text{max}}$, thus, we perform a change of variables to obtain the form of \cref{eq:fgq},
\begin{align}
    \int_a^b f(x) \,dx &= 
    \frac{b-a}{2}&\int_{-1}^1 f\left(\frac{b-a}{2}u +\frac{b+a}{2}\right)\,du.
\end{align}
The same change of variables is performed with the SFB power spectrum $C_\ell(\knl, \knlp)$ in \cref{eq:cl} over a range of $q_\mathrm{max}\gtrsim k_\mathrm{max}$ and $q_\mathrm{min}\lesssim k_\mathrm{min}$ to avoid edge effects. This procedure is powerful in that we can generate a fixed set of $n$ nodes $q_\mathrm{node}$ and weights so that for each $\knl$, we compute $\wnl(\knl, q_\mathrm{node})$. Then, a length $n$ vector with values $\wnl(\knl, q_\mathrm{node})$ contains all of the information needed to calculate $C_\ell(\knl, \knlp)$ provided that we use the same nodes and weights in the integration over $q$. For $q\in (10^{-4}, 0.2)\, h/\mathrm{Mpc}$, we found $n=750$ to yield accurate approximations to the integral \cref{eq:fgq}.

\newpage 
\bibliography{paper}

\end{document}